\documentclass[12pt,a4paper]{article}
\usepackage{latexsym,amsmath,amssymb,verbatim}
\date{}
\numberwithin{equation}{section}

\DeclareMathOperator{\Delb}{\stackrel{\leftrightarrow}{\partial _0}}
\begin{document}
\title{Quantum Supersymmetric Ghosts}
\author{Florin Constantinescu\\ Fachbereich Mathematik \\ Johann Wolfgang Goethe-Universit\"at Frankfurt\\ Robert-Mayer-Strasse 10\\ D 60054
Frankfurt am Main, Germany }
\maketitle

\begin{abstract}
We work out the quantization of the massless vector field by introducing quantum supersymmetric ghosts. We prove positivity in the physical Fock space.
\end{abstract}

\section{The standard Hilbert space of N=1 supersymmetric test functions}

The reason for writting a paper on well known items in the vast literature on supersymmetries is the existence of a positivity (unitarity) structure of the N=1 superspace which was recently put forward in \cite{C1}, and which enables rigorous work on what it could be called the supersymmetric variant of the axiomatic quantum field theory \cite{StreW}. In this section we explain our tools. They are subsequently applied, in an elementary illustrative way, in order to quantize the supersymmetric massles vector field by using free quantum ghosts.\\
We start by considering general supersymmetric (test) functions

\begin{gather}\nonumber 
X(z)=X(x,\theta ,\bar \theta )= \\ \nonumber
=f(x)+\theta \varphi (x) +\bar \theta \bar \chi (x) +\theta ^2m(x)+\bar \theta^2n(x)+ \\ 
\theta \sigma^l\bar \theta v_l(x)+\theta^2\bar \theta \bar \lambda(x)+\bar \theta^2\theta \psi (x)+ \theta^2 \bar \theta^2d(x)
\end{gather}
where $z=(x,\theta ,\bar \theta )$ and the coefficients are functions of certain regularity. Note that the functions (1.1) have the same form as the objects which in supersymmetries are called fields \cite{WB,Sr}. The difference is that in (1.1) we asuume, besides commuting bosonic coefficients, also commuting components of the fermionic coefficients $\varphi (x), \bar \chi (x), \bar \lambda (x), \psi (x) $, i.e. we look at (1.1) as bona fide supersymmetric commuting test functions. The question is if we can induce on these test functions a useful positivity structure (Hilbert space) together with a superdistribution theory such that quantum supersymmetric fields appear as operator-valued superdistributions in such a way that at least a small part of the acchievements in superymmetry can be implemented. First let us remark that contrary to the usual handling of Weyl spinors we need here a van der Waerden calculus with commuting fermionic components (and anticommuting Grassmann variables). It is not difficult to put toghether the rules of this calculus as already done in \cite{C1}. Certainly they differ from the usual rules especially at the point at which we implement complex and Grassmann conjugation \cite{C1}. Working in this framework, let $P_i, i=c,a,t $ be the usual formal disjoint projections on the chiral, antichiral and transversal sectors respectively \cite{WB,Sr}. They satisfy $P_c +P_a +P_t=1 $. Problems with the d'Alembert operator in the denominators are neglected for the moment and will be discussed later. Chiral, antichiral and transversal functions are defined by the conditions
\[ \bar D^{\dot \alpha }X=0, \dot \alpha =1,2;  D^{ \alpha }X=0, \alpha =1,2; D^2 X=\bar D^2 X=0 \]
respectively. On components we have in the chiral case $X_c $

\begin{gather} \nonumber 
\bar \chi =\psi =n=0 , v_l=\partial_l (if)=i\partial_l f ,
  \\ \bar \lambda =-\frac {i}{2}\partial_l \varphi \sigma^l
   =-\frac{i}{2}\bar \sigma^l \partial _l \varphi , 
  d=\frac{1}{4}\square f 
\end{gather}
In the antichiral case $X_a $

\begin{gather} \nonumber
\varphi =\bar \lambda =
m=0,  v_l=\partial_l (-if)=-i\partial_l f , \\ \psi =\frac{i}{2}\sigma^l
\partial_l \bar \chi =\frac{i}{2}\partial _l \bar \chi \bar \sigma^l , d=\frac{1}{4}\square f 
\end{gather}
In the transversal case $X_t $

\begin{gather} \nonumber
\varphi =\bar \lambda =
m=0,  v_l=\partial_l (-if)=-i\partial_l f , \\ \psi =\frac{i}{2}\sigma^l
\partial_l \bar \chi =\frac{i}{2}\partial _l \bar \chi \bar \sigma^l , d=\frac{1}{4}\square f 
\end{gather}
where the vector function $v$ satisfies $\partial _l v^l =0 $
These formulas are very similar to the usual ones but not quite identical to them. The reason is the commuting fermionic convention which in particular modifies the sign in front of $\bar \sigma $.\\
If we restrict on-shell it is clear that there is no nontrivial overlap of sectors. But in the massless case there is a large overlap. A funtion $X$ belongs to this overlap if 

\begin{gather}\nonumber
X(z)=f(x)+\theta \varphi (x)+\bar \theta \bar \chi (x)\pm i\theta \sigma_l \bar \theta \partial_l f(x)
\end{gather}
with
\[\partial_l \varphi \sigma^l =\sigma^l \partial_l \bar \chi =0,\quad \square f=0 \]
Let us return for the moment to the massive case. In \cite{C1} it was shown that the N=1 superspace carries an inherent, invariant Hilbert-Krein structure induced by the positive definite scalar product

\begin{equation}
(X,Y)=<X,\omega Y>=<\omega X,Y>
\end{equation}
where

\begin{equation}
<X,Y>=\int d^8z_1d^8z_2\bar X^T(z_1) K(z_1-z_2) Y(z_2)
\end{equation}
in the notations

\begin{gather}
K (z)=\delta^2 (\theta )\delta^2 (\bar \theta )D^+ (x)\\ 
K_c (z_1 ,z_2 )=P_c K (z_1 -z_2 )\\
K_a (z_1 ,z_2 )=P_a K (z_1 -z_2 )\\
K_T (z_1 ,z_2 )=P_t K (z_1 -z_2 )\\
K_c +K_a +K_t =K
\end{gather}
where $ \delta^2 (\theta^2 )=\theta^2, \delta^2 (\bar \theta^2 )=\bar \theta^2 $ and

\begin{gather}
D^+ (x)=\frac{1}{(2\pi )^2 }\int e^{ipx}d\rho (p),\quad d\rho (p)=\theta (-p_0 )\delta (p^2+m^2 )
\end{gather}
and identification  $ X=\begin{pmatrix}X_c \\X_a \\X_t \end{pmatrix} $  where $X_c =X_1 =P_c X ,X_a =X_2 =P_a X,X_t =X_3 =P_t X$. In (1.5) we have 

\begin{equation}\nonumber
\omega=\begin{pmatrix}1& 0 &0 \\0 &1& 0 \\0& 0 & -1 \end{pmatrix}
\end{equation}
This is a typical Krein structure \cite{Stro} producing as usual the corresponding (maybe not yet physical) Hilbert space of supersymmetric bona fide test functions which will be central for our approach. 
In fact in physics (see later) these two scalar products appear with a minus sign

\begin{equation}
<X,Y>=-\int d^8z_1d^8z_2\bar X^T(z_1) K(z_1-z_2) Y(z_2)
\end{equation}
\begin{equation}
(X,Y)=<X,\omega Y>=<\omega X,Y>
\end{equation}
where

\begin{equation}\nonumber
\omega=\begin{pmatrix}-1& 0 &0 \\0 &-1& 0 \\0& 0 & 1 \end{pmatrix}
\end{equation}
Another form of (1.5) is

\begin{gather} \nonumber
(X,Y)=\int d^8z_1d^8z_2\bar X(z_1)(P_c +P_a -P_t )K_0 (z_1-z_2) Y(z_2)= \\
=<X,(P_c +P_a -P_t )Y>
\end{gather}
Note the surprising minus sign in front of the transversal contribution which turn out to be crucial for the positivity in superspace. A generalization of (1.5), still positive definite, is

\begin{gather} \nonumber
(X,Y)=\int d^8z_1d^8z_2\bar X(z_1)(\lambda_c P_c +\lambda_a P_a +\lambda_t P_t )K_0 (z_1-z_2) Y(z_2)= \\ 
=<X,(\lambda_c P_c +\lambda_a P_a +\lambda_t P_t )Y>
\end{gather}
with positive $\lambda_c ,\lambda_a $ and negative $\lambda_t $.
Note that in the massive case the on-shell restriction is protective against (trivial) zero vectors.\\
In the above Hilbert-Krein framework we can easily define (either by the GNS construction or explicitely in Fock space \cite{C2,C1}) the massive vector field obtaining, at the level of propagators, concordance with the corresponding computation by functional integral methods \cite{WB} and computation on components. However, as to be expected, the functional integral offers no information on positivity. By a corresponding choise of the $\lambda $- constants we can imply several gauges (we are still in the massive case). The choise $ \lambda_c =\lambda_a =\lambda_t =-1$ corresponds to the Feynman gauge. It is not posivide definite. \\
Now let us pass to the, from the physical point of view, more interesting massless case. It turs out that this case is more interesting also from the mathematical point of view. The reason is the d'Alembertian in the denominator of the projection operators. One cheap way to deal with this case would be to simply leave out this operator in the denominators. This has some desadvantages because in this way we apparently loose the sector decomposition of the Hilbert space of supersymmetric functions. Beside that some normalization problems appear. A better (but not more complicated) way which also provides us with the right normalization is to restrict from the beginning the space of general supersymmetric functions to a subspace of it characterized by the following (mild) restrictions on the coefficients \cite{C1} 

\begin{gather}
d(x)=\square D(x) \\
\bar \lambda (x)=\partial_l \Lambda (x)\sigma^l \\
\psi (x)=\sigma^l \partial_l \bar \Psi (x) \\
v_l (x)=\partial_l \rho (x) +\omega_l (x) , \partial_l \omega^l =0 
\end{gather}
where $D(x), \Lambda (x), \bar \Psi (x),\rho (x), \omega (x) $ are arbitrary regular functions. This restriction generates a d'Alembertian which cancells the d'Alembertian in the denominator of the projection operators.
Just in order to have a name we call them restricted (or special) supersymmetric functions. In particular it is easy to see that chiral, antichiral and transversal functions are special supersymmetric. After canceling the d'Alembertian the sector structure of these functions remain untouched. Now consider the (massless) inner products defined  above where the general supersymmetric functions are restricted to  the special ones. The reason of this restriction is that in the restricted Hilbert space to be defined now the projections $P_i ,i=c,a,t $ will be well defined Hilbert space operators. Indeed the overlap between the chiral/antichiral and transversal sectors consists of special supersymmetric functions and what is more interesting all vectors in the overlap are zero vectors (the on shell condition is assumed). The restricted Hilbert space is then obtained as usual by factorization and completion. The restricted Hilbert space will be called simply Hilbert space (in the massless case) and plays the role which we expect from it. \\ Before going over to physical applications, note that in both cases, massive and massles, not only the projections but all formal operators encountered in supersymmetries receive a Hilbert space interpretation (this is not the case in the frame of formal consideration (see examples in \cite{C1}). For instance $i\bar D_{\alpha } $ is the Hilbert space adjoint of $iD_{\alpha } $ and $\bar D^2 $ is the Hilbert space adjoint of $D^2 $. \\
Summarizing, we have obtained in this section the supersymmetric analog of the relativistic Hilbert space $L^2 (R^4,d\mu (x))$ (whose elements are supersymmetric test functions) where $d\mu (x) $ is a Lorentz invariant measure concentrated in the backward light cone. The massles case is different from the massive one. The Hilbert space of supersymmetris is born out of a Krein structure, inherent of the $N=1$ superspace. By dualization we claim here the existence of a supersymmetric Gelfand tripple generalizing the usual relativistic structure (tensor products are defined via the Gelfand tripple). We also claim that the Hilbert space of supersymmetric functions is as usefull in supersymmetry as the ordinary relativistic one in the theory of quantized fields. An example follows in the next section.

\section{The vector field: quantization through quantum ghost fields}

We define the massless vector field as an operator valued superdistribution $V(X)$ generated by the kernel $ -(P_c +P_a +P_t)K(z)=-K(z) $ of the preceeding section on the space of superfunctions. This kernel is suggested also by functional integration methods as well as by a computation on components (in fact from a technical point of view the minus sign in front of the projections is optional). The kernel $K(z)$ is indefinite. A corresponding Fock space representation follows as usual. For specifying the symmetry of the Fock space (and only in this context) we deviate from our convention of commuting components of the Grassmann coeficient functions in (1.1) and adopt for the moment anticommuting fermionic components. In the anticommuting convention we require symmetric Fock spaces. This induces the right symmetry of the Fock space if we return to the working commutative convention. By denoting with 
\[ \Phi =(\Phi^{(0)} ,\Phi^{(1)},\ldots \Phi^{(n)}\ldots ), \quad  \Phi^{(0)}=1, \Phi^{(n)}=\Phi^{(n)}(z_1 ,z_2 \ldots ,z_n ) \]
a general element of the Fock space we have 

\begin{gather}
V(X)=V^{(+)}(X)+V^{(-)}(X)
\end{gather}
with

\begin{gather}
(V^{(-)}(X)\Phi )^{(n)}(w_1 ,\ldots ,w_n )=\sqrt{n+1}(X(w),\Phi^{(n+1)}(w,w_1 ,\ldots ,w_n )) \\ 
(V^{(+)}(X)\Phi )^{(n)}(w_1 ,\ldots ,w_n )=\frac{1}{\sqrt n}\sum_{j=1}^n X(w_j )\Phi^{(n-1)}(w_1,\ldots ,\hat w_j
\ldots ,w_n )
\end{gather}
where $w=(p,\theta ,\bar \theta ) $, $p$ being the momentum variable conjugate to $x$, and $\hat w_j  $ means as usual omission of $w_j $. The annihilation part is denoted by $V^{(-)} (X) $. Note the appearence of the inner product product (.,.) defined in the first section. In the above definition of the massless vector field we assumed $\lambda_i =1,i=c,a,t $ (Feynman) but modulo choosing a gauge we can admit in (.,.) general $\lambda's $. Certainly not all possible choises respect positivity. In \cite{C3} we have quantized the supersymmetric massless vector field following the traditional Gupta-Bleuler quantization in its abstract variat presented in \cite{StroW}. It turns out that the procedure goes through and is by no means more complicated than the usual Gupta-Bleuler quantization. For details the reader can consult \cite{C3}. It is shown there that the restriction from general to special test functions is automatically implemented in the procedure because transversal test functions are special. A translation of the results using supersymmetric creation and distruction operators on the line of \cite{C1} is also possible but not necessary. The celebrated subsidiary condition of Gupta and Bleuler resides in the condition

\begin{gather}
D^2 V^{(-)}(z)=\bar D^2 V^{(-)}(z)=0
\end{gather}
when applied to the Fock space described above. This condition is equivalent with the supression of the chiral and antichiral part in the scalar product. The rest, induced by $-P_t K(z)$, is positive definite (up to the factorization of the zero vectors). If we restrict the supersymmetric test functions  in the definition of the (real) massless vector field by the condition $X=\bar X $ (which we call simply real test functions) then the physical Hilbert space is isomorphic to

\begin{gather}
Ker D^2 /\overline{Im D^2 }=Ker \bar D^2 /\overline {Im \bar D^2 }
\end{gather}
where the bar means closure. \\

In quantum field theory there is an useful exercise which illustrates some aspects of the canonical quantization of abelian gauge theories by using quantum ghosts. It is a particular case of the Kugo-Ojima theory \cite{KO} which applies even for non-abelian theories. Although in this case the ghosts decouple from the electromagnetic potential, it is not entirely without interest \cite{W}. We propose ourself to find its supersymmetric analog. It will face us with the definition of the (free) quantum ghost fields and especially with a rigorous Hilbert space definition of the gauge charge. First let us remember that for the case of chiral fields defined in the chiral/antichiral sector of the Hilbert space we have the following plane wave decomposition \cite{C1}

\begin{gather} 
\phi(x,\theta ,\bar \theta )=\frac {1}{(2\pi )^\frac {3}{2}} \int [a( \bar p,\theta ,\bar \theta )e^{-ipx} + b^+( \bar p,\theta ,\bar \theta )   e^{ipx}]\frac {d^3 \bar p}{\sqrt{2p^0}} \\ 
\bar \phi (x,\theta ,\bar \theta )=\frac {1}{(2\pi )^\frac {3}{2}} \int [b( \bar p,\theta ,\bar \theta )e^{-ipx} + a^+( \bar p,\theta ,\bar \theta )e^{ipx}] \frac {d^3 \bar p}{\sqrt{2p^0}}
\end{gather}
with

\begin{gather}\notag
a(\bar p,\theta ,\bar \theta )=-i\int d^3x e^{ipx} \stackrel{\leftrightarrow}{\partial _0} \phi(x,\theta , \bar \theta ) \\ \notag
b(\bar p,\theta ,\bar \theta )=-i\int d^3x e^{ipx} \Delb \bar \phi(x,\theta ,\bar \theta ) \\ \notag
a^+(\bar p,\theta ,\bar \theta )=-i\int d^3x e^{-ipx} \Delb  \bar \phi(x,\theta ,\bar \theta ) \\ \notag
b^+(\bar p,\theta ,\bar \theta )=-i\int d^3x e^{-ipx} \Delb \phi(x,\theta ,\bar \theta ) \notag
\end{gather}
where as usual 
\[ u\stackrel{\leftrightarrow}{\partial} v=u(\partial v)-(\partial u)v \]
We have

\begin{gather}
[a(\bar p, \theta ,\bar \theta ),b^+(\bar p',\theta ,\bar \theta ')]=0 \\
[a(\bar p, \theta ,\bar \theta ),a^+(\bar p',\theta ,\bar \theta ')]=\frac{\bar D^2 D^2}{16} \delta (\bar p- \bar p')\delta (\theta -\theta ')\delta (\bar \theta -\bar \theta ') \\
[b(\bar p, \theta ,\bar \theta ),b^+(\bar p',\theta ,\bar \theta ')]=\frac{D^2 \bar D^2 }{16} \delta (\bar p - \bar p ')\delta (\theta -\theta ')\delta (\bar \theta -\bar \theta ') \\
[b(\bar p, \theta ,\bar \theta ),a^+(\bar p',\theta ,\bar \theta ')]=0 
\end{gather}
The chiral/antichiral fields above as well as the supersymmetric "creation" and "annihilation" operators are defined in the symmetric chiral/antichiral sector of the Fock space \cite{C1}. Now we modify the commutation relations af the "creation" and "annihilation" operators going from commutators to the anticommutators. Correspondingly we represent them in the chiral/antichiral sector of the antisymmetric Fock space. As usual we use these anticommuting operators in order to define quantum (free) supersymmetric ghost and antighost fields which will be denoted again by $\phi $ and $\bar\phi $. For our restrained purposes we consider only the anticommuting relation of the ghost field $\phi $. It is simply
\[\{\phi (z_1 ),\phi (z_2 )\} =0 \]
for $z_1 ,z_2 $ arbitrary in the superspace. Now we define the supergauge charges by

\begin{gather}
Q=\int D^2 V\stackrel{\leftrightarrow}{\partial_0 }\phi , \quad
\bar Q=\int \bar D^2 V\stackrel{\leftrightarrow}{\partial_0 } \bar \phi
\end{gather}
Correctly speaking $Q,\bar Q $ act in the tensor product of the Fock spaces responsible for $V$ and $\phi ,\bar \phi $. Formally  $Q,\bar Q $ are the antichiral and chiral part of the supersymmetric gauge transformations respectively. According to our restrained purposes we will study only the operator $Q$ given with the help of the ghost field $\phi $. As in the nonsupersymmetric case the gauge charge satisfies the important relation $Q^2 =0 $. The formal proof of this relation (which disregards the tensorial nature of $Q$) goes as usual \cite{KO,S}. It is clear that the physical Hilbert space (restricted to the real test functions above) is given by

\begin{gather}
Ker Q/\overline {Im Q }=Ker D^2 /\overline {Im D^2 }
\end{gather}
The consideration above refer only to the supersymmetric abelin case in which the ghosts totally decouple; it would be intersting to investigate the nonabelian case too, or even more to extend the full Kugo-Ojima theory \cite{KO} to the supersymmetric case. It is clear that the standard Hilbert-Krein structure of superymmetries \cite{C1} has to be used in this context.


\begin{thebibliography}{99}

\bibitem {C1} F. Constantinescu, Lett. Math. Phys, 62(2002), 111, hep-th/0404182, hep-th/0408195
\bibitem {StreW} R.F. Streater, A.S. Wightman, PCT, Spin and Statistics and All That, Benjamin, 1964
\bibitem {Stro} F. Strocchi, Selected topics on the General Properties of Quantum Field theory, World Scientific, 1993 
\bibitem {WB} J. Wess, J.Bagger, Supersymmetry and Supergravity, 2nd edition, Princeton University Press, 1992
\bibitem {Sr} P.P. Srivastava, Supersymmetry, Superfields and Supergravity: An Introduction, IOP Publishing, Adam Hilger, Bristol, 1986
\bibitem {C2} F. Constantinescu, hep-th/0305143, hep-th/0306019
\bibitem {C3} F. Constantinescu, hep-th/0408046
\bibitem {StroW} F. Strocchi, A.S. Wightman, Jornal Math.Phys., 12(1974), 2198
\bibitem {KO} T. Kugo, I. Ojima, Progress of Theoretical Physics, Supplement, 66(1979), 1
\bibitem {W} S. Weinberg, The Quantum Theory of Fields  vol. 2, Cambridge University Press, 1996
\bibitem {S} G. Scharf, Quantum Gauge Theories, J. Wiley, 2001





\end{thebibliography}
\end{document}